\documentclass[11pt]{article}

\usepackage{amsmath,amsthm,amssymb,cite,enumerate,graphicx,epsfig} % when quoting [1,2,3,4] this gives simply [1-4]

\usepackage[usenames]{color}
\usepackage{bm}
\usepackage{authblk} %for authors with multiple affiliations

\def \BEA { \begin{eqnarray}}
\def \EEA {\end{eqnarray}}
\def \BE {\begin{equation}}
\def \EE {\end{equation}}
\def\d{\mathrm{d}}

\def \bF {\mbox{\boldmath{$F$}}}

\def \WDS #1 {\mbox{$\Phi_{#1}^{S}$}}
\def \WDA #1 {\mbox{$\Phi_{#1}^{A}$}}
\def \WD #1 {\mbox{$\Phi_{#1}$}}

\def \mi {\stackrel{i}{m}}
\def \mj {\stackrel{j}{m}}
\def \mk {\stackrel{k}{m}}
\def \mr {\stackrel{r}{m}}
\def \ms {\stackrel{s}{m}}
\def \mz {\stackrel{z}{m}}
\def \mq {\stackrel{q}{m}}
\def \mo {\stackrel{o}{m}}
\def \mD {\stackrel{2}{m}}
\def \mT {\stackrel{3}{m}}
\def \mC {\stackrel{4}{m}}

\def \mio #1 {\mi_{#1}\ ^{  \! \! \! \! 0}} 
\def \mjo #1 {\mj_{#1}\ ^{  \! \! \! \! 0}} 
\def \mko #1 {\mk_{#1}\ ^{  \! \! \! \! 0}} 
\def \mro #1 {\mr_{#1}\ ^{  \! \! \! \! 0}} 
\def \mso #1 {\ms_{#1}\ ^{  \! \! \! \! 0}} 
\def \mpo #1 {\mp_{#1}\ ^{  \! \! \! \! 0}} 
\def \mzo #1 {\mz_{#1}\ ^{  \! \! \! \! 0}} 
\def \mqo #1 {\mq_{#1}\ ^{  \! \! \! \! 0}} 
\def \moo #1 {\mo_{#1}\ ^{  \! \! \! \! 0}} 
\def \mDo #1 {\mD_{#1}\ ^{  \! \! \! \! 0}} 
\def \mTo #1 {\mT_{#1}\ ^{  \! \! \! \! 0}} 
\def \mCo #1 {\mC_{#1}\ ^{  \! \! \! \! 0}} 

\def \miJ #1 {\mi_{#1}\ ^{  \! \! \! \! (1)}} 
\def \mjJ #1 {\mj_{#1}\ ^{  \! \! \! \! (1)}} 
\def \mkJ #1 {\mk_{#1}\ ^{  \! \! \! \! (1)}} 
\def \mrJ #1 {\mr_{#1}\ ^{  \! \! \! \! (1)}}

\def \bk {\mbox{\boldmath{$k$}}}

\def \hbm #1 {\mbox{\boldmath{$\hat m^{(#1)}$}}}

\def \bk {\mbox{\boldmath{$k$}}}

\newcommand{\be}{\begin{equation}}
\newcommand{\ee}{\end{equation}}
\newcommand{\beqn}{\begin{eqnarray}}
\newcommand{\eeqn}{\end{eqnarray}}
\newcommand{\pa}{\partial}
\newcommand{\ba}{\begin{array}}
\newcommand{\ea}{\end{array}}
\newcommand{\pp}{{\it pp\,}-}

\def \BEAH {\begin{eqnarray*}}
\def \EEAH {\end{eqnarray*}}
\def \BEA {\begin{eqnarray}}
\def \EEA {\end{eqnarray}}
\def \BDM {\begin{displaymath}}
\def \EDM {\end{displaymath}}

\begin{document}

\title{A note on Kundt spacetimes of type N with a cosmological constant}

\author[1,2]{Marcello Ortaggio\thanks{ortaggio(at)math(dot)cas(dot)cz}}

\affil[1]{Institute of Mathematics of the Czech Academy of Sciences, \newline \v Zitn\' a 25, 115 67 Prague 1, Czech Republic}
\affil[2]{Instituto de Ciencias F\'{\i}sicas y Matem\'aticas, Universidad Austral de Chile, \newline Edificio Emilio Pugin, cuarto piso, Campus Isla Teja, Valdivia, Chile}

\date{\today}

\maketitle

\begin{abstract}

In recent literature there appeared conflicting claims about whether the Ozsv\'ath-Robinson-R\'ozga family of type N electrovac spacetimes  of the Kundt class with $\Lambda$ is complete. We show that indeed it is.

\end{abstract}

\bigskip

\section{Introduction}

\label{sec_intro}

Kundt spacetimes are defined by the existence of a congruence of null geodesics with zero expansion, twist and shear. In Kundt coordinates, the line-element takes the form \cite{Kundt61} (cf. also the reviews \cite{Stephanibook,GriPodbook})
\be
 \d s^2=2P^{-2}\d\zeta\d\bar\zeta-2\d u\left(\d r+W\d\zeta+\bar W\d\bar\zeta+H\d u\right) ,
\label{kundt}
\ee
where the metric functions $P$ and $H$ are real, with $P_{,r}=0$, and $W$ is complex. The Kundt null direction is along $\pa_r$. Kundt spacetimes which satisfy Einstein's equations with aligned pure radiation and which are of Petrov type III and N were obtained already in \cite{Kundt61} and consists of two invariant subclasses, i.e., \pp waves and Kundt waves \cite{Stephanibook,GriPodbook}. Adding a cosmological constant makes the field equations more complicated, but also gives rise to a richer class of solutions. The first systematic study of Kundt Einstein spacetimes of type N was performed in \cite{GarPle81}. An invariant classification of all such solution was subsequently presented in \cite{OzsRobRoz85} (also allowing for an aligned null electromagnetic field), were it was additionally pointed out that for $\Lambda<0$ there exist some new solutions in addition to those of \cite{GarPle81} (see also \cite{BicPod99I,GriPodbook} for related comments). More recently, the results of \cite{OzsRobRoz85} have been generalized to the type III \cite{GriDocPod04}. As a side result, Ref.~\cite{GriDocPod04} also claims to present a new family of type N solutions, not identified in the previous literature, while the classification given in \cite{OzsRobRoz85} was previously considered to be complete. Given the importance of exact solutions describing gravitational waves in the presence of a cosmological constant, we believe that this discrepancy should be clarified. This is of interest not only in Einstein's theory but in virtually any metric theory of gravity, since the solutions of \cite{OzsRobRoz85} are universal spacetimes \cite{HerPraPra14}. It is the purpose of the present note to show that the type N solutions of \cite{GriDocPod04} can in fact be reduced to those of \cite{OzsRobRoz85}. All Kundt spacetime of type N with a cosmological constant and an aligned Maxwell field can thus be locally written in the form given in \cite{OzsRobRoz85}.

\section{Analysis of the type N solutions}

The most general Kundt spacetime of Petrov III which solves Einstein's equations with a cosmological constant $\Lambda$ and aligned pure radiation\footnote{In an adapted Newman-Penrose null tetrad, this means $\Psi_0=\Psi_1=\Psi_2=0=\Phi_{00}=\Phi_{01}=\Phi_{02}=\Phi_{11}=\Phi_{12}$.} can be written as \eqref{kundt} with \cite{GriDocPod04}
\beqn
  & & W=2r\phi_{,\zeta}+\psi_{,\zeta} , \\
	& & H=-r^2ke^{2\phi}-r\left[P^2(\phi_{,\bar\zeta}\psi_{,\zeta}+\phi_{,\zeta}\bar\psi_{,\bar\zeta})+2\lambda(\psi+\bar\psi)\right]+H^{0} , \label{H} \\
	& & P=1+\lambda\zeta\bar\zeta , \qquad Q=a\left(1-\lambda\zeta\bar\zeta\right)+\bar b\zeta+b\bar\zeta , \qquad \phi=\ln\left|\frac{P}{Q}\right| , \qquad k=\lambda a^2+b\bar b , \qquad \lambda=\frac{1}{6}\Lambda , \label{PQ}
\eeqn
where $\psi=\psi(u,\zeta)$ and $H^{0}=H^{0}(u,\zeta,\bar\zeta)$ are arbitrary functions of their arguments, while $a$ and $b$ are constants\footnote{To be precise, in \cite{GriDocPod04} it was shown that coordinates can be chosen such that the spin coefficient $\bar\tau=P\phi_{,\zeta}$ does not depend on $u$. This allows for a factorized $u$-dependence of $Q$, i.e., $Q=Q_1(u)Q_2(\zeta,\bar\zeta)$, which however does not enter the metric and can thus be dropped.}  (with $H^{0}$ and $a$ real, $\psi$ and $b$ complex, $a$ and $b$ not both vanishing).
Similarly as in \cite{OzsRobRoz85}, canonical forms of the metric correspond to various specific choices of $a$ and $b$, which depend on the (invariant) sign of $k$ \cite{GriDocPod04} (however, this is not needed in the following discussion).
For later computations, it is useful to observe that $ke^{2\phi}=P^2\phi_{\zeta}\phi_{\bar\zeta}+\lambda$. Based on \cite{OrtPra18} and on the $\Phi_{22}$ component given in \cite{GriDocPod04}, we further observe that the matter content of these spacetimes can be interpreted as an aligned null electromagnetic field $\bF=\d u\wedge\left[f(u,\zeta)\d\zeta+\bar f(u,\bar\zeta)\d\bar\zeta\right]$ provided $H^0$ satisfies
\beqn
	& & H^0_{,\zeta\bar\zeta}+\phi_{,\bar\zeta}H^0_{,\zeta}+\phi_{,\zeta}H^0_{,\bar\zeta}+2P^{-2}(ke^{2\phi}+\lambda)H^0 \nonumber \\ 
	& & \qquad\qquad\qquad {}+\phi_{,\bar\zeta}\bar\psi_{,\bar\zeta}\left(P^{2}\psi_{,\zeta}\right)_{,\zeta}+\phi_{,\zeta}\psi_{,\zeta}\left(P^{2}\bar\psi_{,\bar\zeta}\right)_{,\bar\zeta}+2(ke^{2\phi}+2\lambda)\psi_{,\zeta}\bar\psi_{,\bar\zeta}=2f\bar f.
\eeqn
Setting $f=0$ gives rise to vacuum solutions.

Requiring the Petrov type to be N (i.e., $\Psi_3=0$) gives \cite{GriDocPod04} 
\be
  \psi_{,\zeta}=\frac{c}{PQ\phi_{,\bar\zeta}}=\frac{c}{-b+2a\lambda\zeta+\bar b\lambda\zeta^2} ,
\ee 
where $c$ is an arbitrary complex function of $u$. This also means that in \eqref{H} one has $P^2(\phi_{,\bar\zeta}\psi_{,\zeta}+\phi_{,\zeta}\bar\psi_{,\bar\zeta})=e^\phi(c+\bar c)$. We now want to show that these type N solutions can be transformed into those found in \cite{OzsRobRoz85}. Since there is no disagreement in the literature concerning the $\lambda=0$ case, in the rest of this section we will assume $\lambda\neq0$.

It is convenient to proceed by introducing the coordinates of \cite{OzsRobRoz85}, defined by
\be
 r=e^{-2\phi}v ,
\ee
such that
\be
 \d s^2=2P^{-2}\d\zeta\d\bar\zeta-2e^{-2\phi}\d u\left[\d v+e^{2\phi}\psi_{,\zeta}\d\zeta+e^{2\phi}\bar \psi_{,\bar \zeta}\d\bar\zeta+e^{2\phi}H\d u\right] ,
\label{ORR}
\ee
with \eqref{PQ} and 
\be
 e^{2\phi}H=-kv^2-v\left[e^\phi(c+\bar c)+2\lambda(\psi+\bar\psi)\right]+e^{2\phi}H^{0} .
 \label{H_2}
\ee

Let us first discuss the case when $a\neq0$ in \eqref{PQ}. By defining
\be
 \zeta=e^{i\theta(u)} z ,
 \label{transf1}
\ee
where $\theta$ is real, one obtains
\be
 \d s^2=2P^{-2}\d z\d\bar z-2e^{-2\phi}\d u\left[\d v+\bar Z\d z+Z\d\bar z+S\d u\right] ,
\label{ORR2}
\ee
with 
\beqn
& & \bar Z=e^{2\phi}\psi_{,z}+iQ^{-2}\dot\theta\bar z , \qquad \psi_{,z}=\frac{c}{PQ\phi_{,\bar z}} \label{Z'} \\
 & & S=-kv^2-v\left[e^\phi(c+\bar c)+2\lambda(\psi+\bar\psi)\right]+e^{2\phi}\left[H^{0}+i\dot\theta(z\psi_{,z}-\bar z\bar\psi_{,\bar z})-P^{-2}\dot\theta^2 z\bar z\right] , \\
& & P=1+\lambda z\bar z , \qquad Q=a\left(1-\lambda z\bar z\right)+\bar b e^{i\theta}z+b e^{-i\theta}\bar z , \label{PQ2}
\eeqn
$\phi$ and $k$ as in \eqref{PQ}, and a dot denoting differentiation w.r.t. $u$. Since $\bar Z_{,\bar z}-Z_{,z}=2PQ^{-3}(c-\bar c+ia\dot\theta)$, following \cite{OzsRobRoz85} we note that one can set 
\be
 \bar Z_{,\bar z}-Z_{,z}=0 ,
 \label{dZ}
\ee
by choosing $\theta(u)$ such that
\be
 i\dot\theta=-\frac{c-\bar c}{a} \qquad (a\neq 0) .
 \label{theta}
\ee
Then, by transforming
\be
 v=v'+\chi(u,z,\bar z) ,
\ee
thanks to \eqref{dZ} we can set simultaneously $\bar Z=0=Z$, provided $\chi$ is a real solution of
\be
 \chi_{,z}+\bar Z=0 .
 \label{chi}
\ee

We thus arrive at the line-element
\be
 \d s^2=2P^{-2}\d z\d\bar z-2e^{-2\phi}\d u\left[\d v'+S'\d u\right] ,
\label{ORR3}
\ee
with 
\beqn
 & & S'= -kv'^2-v'\left[e^\phi(c+\bar c)+2\lambda(\psi+\bar\psi)+2k\chi\right]+\frac{1}{2}e^{\phi}{\cal H}(u,z,\bar z) , 
\eeqn
where we have defined (recall that, at this stage, $H^{0}(u,z,\bar z)$ is an arbitrary function)
\be
  \frac{1}{2}e^{\phi}{\cal H}=e^{2\phi}\left[H^{0}+i\dot\theta(z\psi_{,z}-\bar z\bar\psi_{,\bar z})-P^{-2}\dot\theta^2 z\bar z\right]-k\chi^2-\chi\left[e^\phi(c+\bar c)+2\lambda(\psi+\bar\psi)\right]+\chi_{,u} ,
\ee
and with \eqref{Z'}--\eqref{PQ2}, \eqref{theta} and \eqref{chi}. Similarly as in \cite{OzsRobRoz85}, one can notice that $S'_{,v'z}=(\ln|Q|)_{,uz}$ and therefore, by exploiting a freedom of redefining $u\mapsto U(u)$ (and, if necessary, $v'\mapsto -v'$),  without loosing generality one can set in \eqref{ORR3}
\beqn
 S'= -kv'^2+v'(\ln|Q|)_{,u}+\frac{1}{2}e^{\phi}{\cal H} .
 \label{S_red}
\eeqn
The function $Q$ is still of the form \eqref{PQ2}, but now $a$ and $b$ are generically functions of $u$ ($\theta(u)$ can therefore be reabsorbed into $b$). Metric \eqref{ORR3} with \eqref{S_red} is precisely of the form obtained in \cite{OzsRobRoz85} (see eq.~(4.36) therein).

In the case $a=0$ (which requires $b\neq0$ and thus $k>0$), the coordinate transformation~\eqref{transf1} cannot be used to arrive at \eqref{dZ}. Nevertheless, one can employ a transformation of the form (a special case of those discussed in \cite{OzsRobRoz85})
\be
 \zeta=\frac{e^{i\theta(u)}z-\frac{\gamma}{\bar b}e^{-i\theta(u)}}{\frac{\gamma}{b}\lambda e^{i\theta(u)}z+e^{-i\theta(u)}} , 
 \label{transf2}
\ee
where the dimensionless real constant $\gamma$ is non-zero and such that $b\bar b+\lambda\gamma^2\neq 0$. The condition \eqref{dZ} is now satisfied provided
\be
 i\dot\theta=(c-\bar c)\frac{\gamma^2\lambda+b\bar b}{4\gamma b\bar b}  \qquad (a=0) .
 \label{theta2}
\ee
The rest of the argument is the same as in the case $a\neq0$, which leads again to a metric of the form \eqref{ORR3} with \eqref{S_red}, which is contained in the solutions of \cite{OzsRobRoz85}. We have thus shown that the type N solutions of \cite{GriDocPod04} are contained in the family obtained in \cite{OzsRobRoz85}. As a consequence \cite{HerPraPra14}, the latter contains all universal spacetime of type N in four dimensions.

\section{Summary of the classification}

To conclude, it may be useful to briefly summarize the results of \cite{OzsRobRoz85}. The most general type N Kundt spacetime satisfying Einstein's equations in the presence of aligned pure radiation and an arbitrary cosmological constant can be locally written as (dropping the primes in \eqref{ORR3} and \eqref{S_red} and recalling $e^{2\phi}=P^2Q^{-2}$)
\beqn
 & & \d s^2=\frac{2}{P^{2}}\d z\d\bar z-2\frac{Q^2}{P^{2}}\d u\left[\d v+\left(-kv^2+v\frac{Q_{,u}}{Q}\right)\d u\right]-\frac{Q}{P}{\cal H}\d u^2 , \label{ORR_final} \\
 & & P=1+\lambda z\bar z , \qquad Q_{,v}=0, \qquad {\cal H}_{,v}=0 ,
 \label{S_final}
\eeqn
where $\bk=\pa_v$ defines the principal null direction of the Weyl tensor and the function ${\cal H}(u,z,\bar z)$ is arbitrary. Using a coordinate freedom (cf. \cite{OzsRobRoz85,BicPod99I} for details), $Q$ and $k$ can always be reduced to the following canonical forms, in the invariantly defined cases $k>0$, $k<0$ and $k=0$, respectively:
\begin{enumerate}
 
	\item\label{k+} $k=+1$, $Q=z+\bar z$, $\lambda$ can have any sign. Setting $\lambda=0$ gives rise to Kundt waves \cite{Kundt61}, while the $\lambda\neq 0$ solutions, first obtained in \cite{GarPle81} (cf. also \cite{BicPod99I}), are often referred to as ``generalized Kundt waves''.
	
	\item $k=\lambda<0$, $Q=1-\lambda z\bar z$. This branch was discovered in \cite{OzsRobRoz85} (to be precise, it also exists for $k=\lambda>0$, however this case can be omitted, since it can be transformed into case~\ref{k+}, cf.~\cite{BicPod99I}).
	
		\item $k=0$, $\lambda\le 0$, $Q=(1+\sqrt{-\lambda}e^{i\alpha(u)}z)(1+\sqrt{-\lambda}e^{-i\alpha(u)}\bar z)$. Also this branch was discovered in \cite{OzsRobRoz85}. The special subcase $\alpha_{,u}=0$ was first found in \cite{Siklos85} using different coordinates (as discussed in \cite{Podolsky98sik,BicPod99I}) and admits $\bk$ as a Killing vector field. For $\lambda=0$ one recovers \pp waves.

\end{enumerate}

Some comments on the character of the singularities of these solutions can be found in \cite{GriDocPod04}.

In all cases, the aligned pure radiation can be interpreted as a null electromagnetic field $\bF=\d u\wedge\left[f(u,z)\d z+\bar f(u,\bar z)\d\bar z\right]$ provided ${\cal H}$ satisfies \cite{OzsRobRoz85}
\be
 {\cal H}_{,z\bar z}+2\lambda P^{-2}{\cal H}=4PQ^{-1}f\bar f.
 \label{electrovac}
\ee
Vacuum solutions are obtained by integrating \eqref{electrovac} with $f=0$, which gives \cite{OzsRobRoz85}
\be
 {\cal H}=h_{,z}+\bar h_{,\bar z}-2\lambda P^{-1}(\bar z h+z\bar h) ,
 \label{vacuum}
\ee
where $h=h(u,z)$.

For ${\cal H}=0$, metric~\eqref{ORR_final} describes a space of constant curvature (more generally, this occurs for \eqref{vacuum} with $h=a_0(u)+a_1(u)z+a_2(u)z^2$, with complex $a_i$)  \cite{OzsRobRoz85,BicPod99I}, which shows that all spacetimes \eqref{ORR_final} (independently of the field equations) belong to the (A)dS-Kerr-Schild class.

\section*{Acknowledgments}

I am grateful to Ji\v{r}\'{\i} Podolsk\'y for reading the manuscript. This work has been supported by research plan RVO: 67985840. The author acknowledges support from the
Albert Einstein Center for Gravitation and Astrophysics, Czech Science Foundation GACR 14-37086G. The author's stay at Instituto de Ciencias F\'{\i}sicas y Matem\'aticas, Universidad Austral de Chile has been supported by CONICYT PAI ATRACCI{\'O}N DE CAPITAL HUMANO AVANZADO DEL 
 EXTRANJERO Folio 80150028.

%%\bibliographystyle{JHEP}
%\bibliographystyle{unsrt}
%\bibliography{bibl}

\end{document}